# Symmetry and Temperature Dependence of the Order Parameter in $MgB_2$ from Point Contact Measurements

A. Kohen and G. Deutscher

School of physics and Astronomy, Raymond and Beverly Sackler Faculty of Exact Sciences, Tel Aviv University, 69978 Tel Aviv, Israel

**Abstract**

We have performed differential conductance versus voltage measurements of Au/$MgB_2$ point contacts. We find that the dominant component in the conductance is due to Andreev reflection. The results are fitted to the theoretical model of BTK for an s-wave symmetry from which we extract the value of the order parameter (Δ) and its temperature dependence. From our results we also obtain a lower experimental bound on the Fermi velocity in $MgB_2$.

Recently superconductivity was discovered in magnesium diboride by Nagamatsu et al. (1) with a critical temperature $T_c$ of 39 K. The discovery was followed by weak link measurements in the tunneling regime. Rubio Bollinger et al. (2) used a scanning tunneling microscope (STM) to measure tunneling into small grains of $MgB_2$ embedded in a gold matrix. The measurement was done at a temperature of 2.5 K. A good fit was found to the BCS model with a s-wave symmetry gap (Δ=2 meV). A. Sharoni et al. (3) used STM to measure a bulk sample of $MgB_2$ at T=4.2

K. They also found a good fit to a BCS model with an isotropic order parameter of a larger amplitude ($\Delta$=5-7 meV). We report here on the temperature dependence of point contact measurements on $MgB_2$ in the Sharvin limit (4). From these measurements we extract the temperature dependence of the order parameter and calculate a lower bound for the Fermi velocity in $MgB_2$. The resulting spectra can be fitted according to the BTK (5) formalism with an s-wave symmetry order parameter, with amplitudes in the range of 3-4 meV at $T \ll T_c$. The barrier parameter $Z=\frac{H}{\hbar \bar{v}_F}$ (where H is the height of the potential between the normal metal and the superconductor) values obtained were between 0.57 to 0.9, indicating a dominant Andreev reflection (6) component in the conductance. Andreev reflection is a unique property of a superconducting material, in which a phase coherent state consisting of Cooper pairs is formed below $T_c$. This reflection occurs at the interface between a normal metal and a superconductor. An electron approaching the superconductor from the normal metal with energy smaller than the energy gap of the superconductor cannot enter as a quasiparticle into the superconducting condensate. Instead the electron is reflected as a hole and a Cooper pair is added to the condensate. This process results in an increase of the conductivity of the contact for voltages smaller then $\Delta / e$ (where e is the electron charge). This is different from the case of a tunnel junction, in which one measures a decrease in the conductance below $\Delta$, resulting from a decrease in the density of states of quasiparticles in the superconductor.

The $MgB_2$ sample, that we measured is of the same source as that used in Ref. (3) and was prepared following the procedure reported in Ref. (7). The superconducting transition found by a magnetization measurement gave $T_c$=39 K. (3). Details on sample preparation and characterization can be found in Ref. (3). The point contacts were obtained using an Au tip mounted on a differential screw. Details on the technique can be found in Achsaf et al. (8). Prior to the measurement the sample was polished with a silicone carbide paper (2500 grit). I (V) characteristics of the contacts were measured digitally and differentiated numerically using a computer program. Each measurement comprises of two successive cycles in order to check for the absence of heating hysteresis effects. The conductance curves are symmetrical with respect to voltage. All figures show the spectra after normalization, the conductance data being divided by the value of the conductance at $V>\Delta$, where it reaches a constant value.

In figure 1 we show the differential conductance of the highest resistance contact R (V=25 mV)= 45 Ohm, at a temperature of 4.2 K. The data was fitted using an s-wave symmetry order parameter with $\Delta$=4 meV, Z=0.9 and a smearing factor $\Gamma$=2 meV. Figure 2 shows the data for a contact with a resistance of R (V=25 mV)= 24 Ohm at a temperature of 4.2 K. Fitting parameters are $\Delta$=3.8 meV, Z=0.75 and $\Gamma$=1.5 meV.

In figure 3a we show the characteristics of our lowest Z contact, measured at 7.4 K. The contact resistance was R (V=25 mV)=9 Ohm. The data was fitted with s-wave symmetry $\Delta$=3 meV, Z=0.57 and $\Gamma$=0.75 meV. From this Z value we can calculate an upper bound for the ratio of the Fermi velocities of Au and $MgB_2$. If we use the relation $Z_{eff}=[Z^2+(1-r)^2/4r]^{1/2}$ of Ref. (2), where r is the ratio of the Fermi velocities, we find that r<3, which gives for the Fermi velocity of $MgB_2$ a lower bound of $4.7x10^7$ cm/sec (where we have used $1.4x10^8$ cm/sec as the Fermi velocity of Au (9)). This in agreement with the value of the average Fermi velocity of $4.8x10^7$ cm/sec calculated by Kurtus et al. (10) At voltages above 6 mV the BTK theory fails however to explain the data. First the measured conductance rises above the BTK fit until it reaches a maximum separation around 8.6 mV, then it crosses the fit line and continues below it until a maximum separation at around 15 mV. The data and the fit join around 20 mV. This behavior is seen both for negative and for positive bias. The misfit may be due to the bosons mediating the attractive el-el interaction. We subsequently measured the same contact at different temperatures; 10K, 14 K and 25 K. This is shown in Fig. 3b, 3c and 3d. We find that the resistance at high voltage is constant at the different temperatures. We used the same Z and the same Γ, as for the 7.4 K measurement and changed only Δ and T, to fit the data. From this procedure we were able to extract Δ as a function of temperature temperature in fig4. We were able to fit the data to the BCS prediction using Δ(0)=3 meV and $T_c$=29 K. This $T_c$ is lower then the bulk critical temperature of 39 K. However if we assume that the highest gap value we measured of 4 meV corresponds to the bulk $T_c$ and that Δ is proportional to a local $T_c$ we get $T_c$(Δ=3 meV)=(3/4)*39=29.3 K. This is then in agreement with our fitted value. In any case, the value of $T_c$ predicted by the weak coupling limit, $\Delta(0)/k_BT=1.76$, for Δ(0)=3 meV is 19.8 K, while our Data shows that $T_c$ of the contact is definitely above 25 K. This gives an upper limit to the ratio $\Delta(0)/k_BT_c$ of 1.4, lower than the BCS weak coupling value of 1.75. We obtain for $\Delta(0)/k_BT_c$ the same value of 1.4 if we use our highest measured value of Δ=4 meV and $T_c$=39

K. Using the BCS expression $\xi_0 = \frac{\hbar v_F}{\pi\Delta}$, and our lower bound for $v_F$=4.7x10$^7$ cm/sec, gives $\xi_0 \cong$ 250 A$^o$. This value is smaller than the mean free path value of 600 A$^o$ given by Ref. (11), thus we find that $MgB_2$ is intrinsically in the clean limit, since the value of $\xi_0$ that we calculated is independent of the mean free path. The value of the mean free path is also larger than the size of the point contact a $\cong$ 20 to 40 A$^o$, which we calculate from the measured contact resistance and the fitted Z value. (Using the relations $R_n=R_0(1+Z^2)$ and $R_0=\rho l/4a^2$(5) and the value for ρ from Ref.(11)). Our contacts are thus in the Sharvin limit. (a<<l)

In conclusion we showed data of low Z, point contact measurements on $MgB_2$. The data was fitted using the BTK model for an s-wave symmetry order parameter. The data has a dominant component of Andreev reflection, which is a sign of a phase coherent state formed by the electrons. From the Z value of the fit, we calculated a lower bound of 4.7x10$^7$ cm/sec for the Fermi velocity $MgB_2$. The temperature dependence of the order parameter amplitude is consistent with the BCS prediction. However an upper limit on the ratio $2\Delta(0)/k_BT$ of 2.8 is found which is smaller than the BCS weak limit prediction.

We would like to thank I. Felner for supplying the $MgB_2$ sample and I.Felner O. Millo and A. Sharoni for useful discussions. This work was supported in part by the Heinrich Hertz-Minerva Center for High Temperature Superconductivity, by a grant from DARPA and ONR, and by the Oren Family Chair of Experimental Solid State physics.

References:

1. J. Nagamatsu et al. (sumitted to Nature).
2. G. Rubio-Bollinger et al. Cond-mat/0102242(unpublished)
3. A. Sharoni, I. Felner and O. Milo, cond-mat, 0102325(unpublished)
4. Yu . V. Sharvin , Zh. Eksp. Teor. Fiz. **48**, 984 (1965).
5. G.E. Blonder, M. Tinkham, and T.M. Klapwijk, Phys. Rev. B **25**, 4515 (1982).
6. A. F. Andreev, Zh. Eksp. Teor. Fiz. **46**, 1823 (1964).
7 -mat,0101463(2001)
8. N. Achsaf, G.Deutscher , A. Revcolevschi and M. Okuya in Coherence in High Temperature Superconductivity, edited by G. Deutscher and A. Revcolevschi (world Scientific, 1996).
9. N.W. Ashcroft and N.D. Mermin, Solid State Physics (Saunders College, HRW International Edition 1987), p.38.
10. J. Kurtos et al. , cond-mat/0101446 (unpublsihed)
11. P. C. Canfield et al. , Phys Rev Lett. **86** , 2423 (2001)

Figure Captions:

Figure 1.

Normalized conductance versus voltage of Au/$MgB_2$ contact measured at 4.2 K. R (V=25 mV)= 45 Ohm. (Circles). BTK fit: $\Delta$=4 meV, Z=0.9, T=4.2K and $\Gamma$=2.0 meV (Line).

Figure 2

Normalized conductance versus voltage of Au/$MgB_2$ contact measured at 4.2 K. R(V=25 mV)= 25 Ohm. (Circles) . BTK fit: $\Delta$=3.8 meV, Z=0.75, T=4.2K and $\Gamma$=1.5 meV (Line).

Figure 3

Normalized conductance versus voltage of Au/$MgB_2$ contact measured as a function of temperature. R(V=25 mV)= 9 Ohm. (Circles). BTK fit: Z=0.57 and $\Gamma$=0.75 meV (Line). (a) T=7.4 K $\Delta$=3 meV, (b) T=10K $\Delta$=3 meV, (c) T=14K $\Delta$=2.7 meV (d) T=25K $\Delta$=1.8 meV

Figure 4

Amplitude of the order parameter ($\Delta$) as a function of temperature. Data of the contact presented in Fig. 3. (Squares). BCS fit with $\Delta$(T=0)=3 meV and $T_c$=29 K (Line)

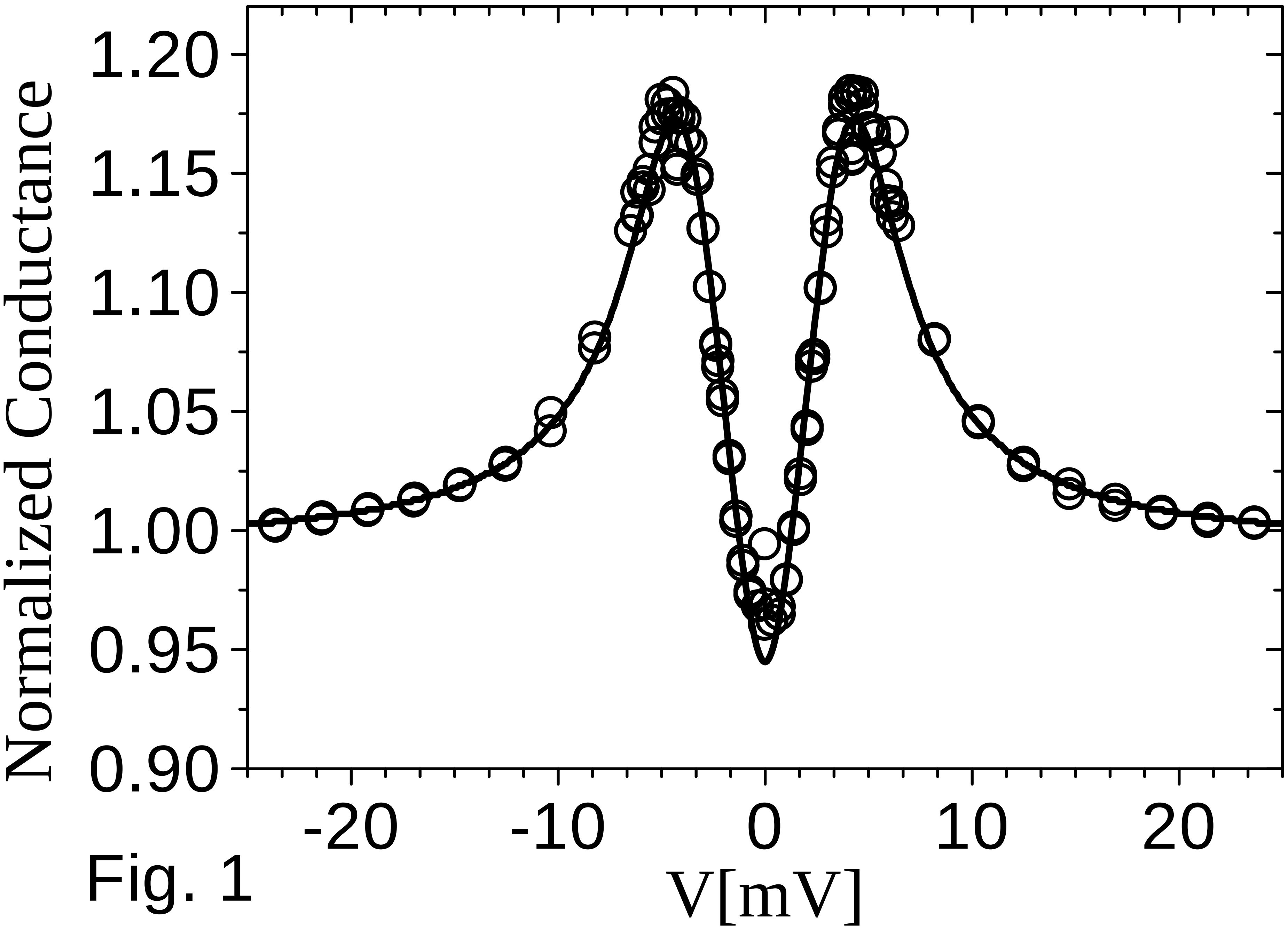

Fig. 1

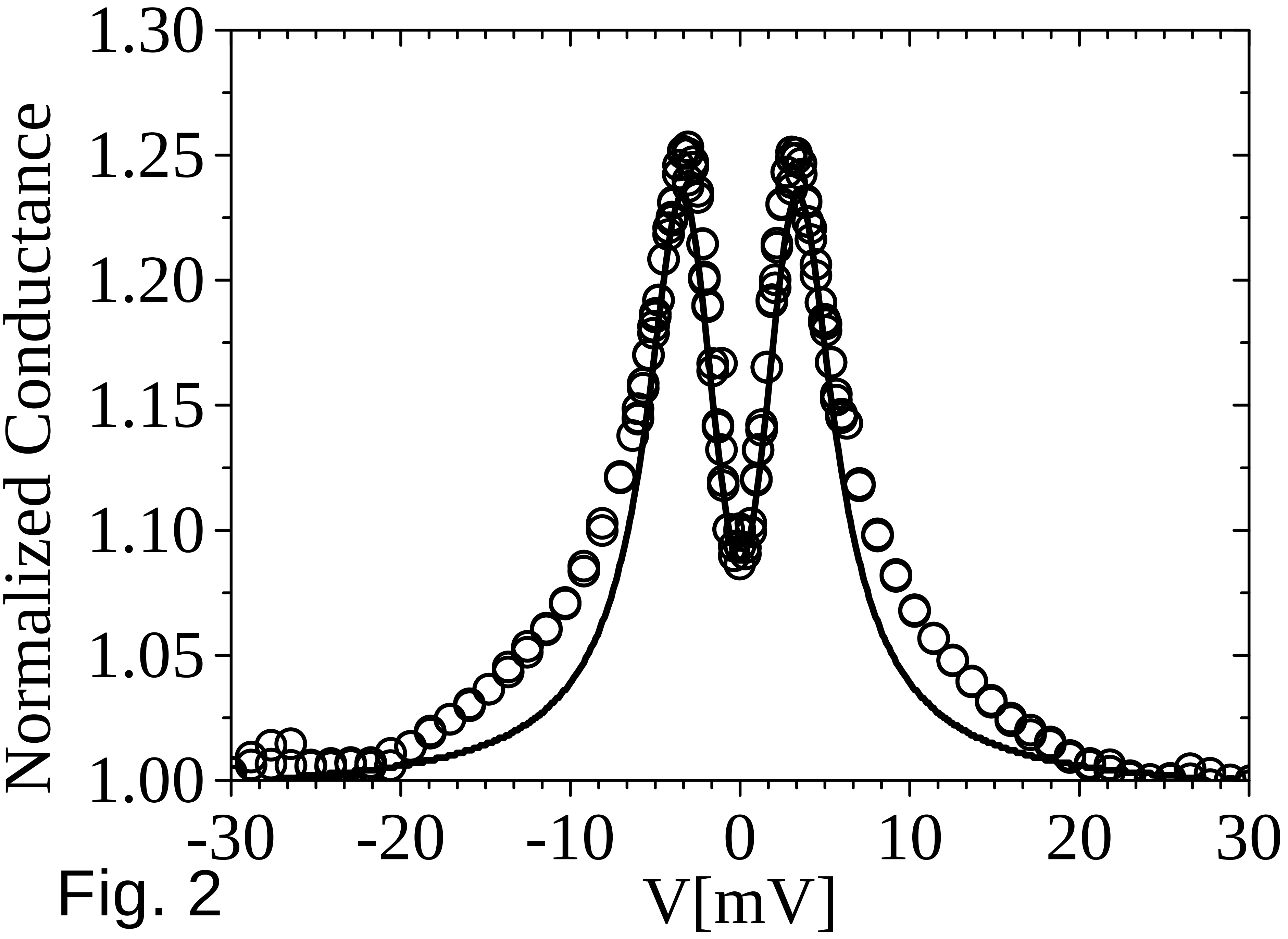

Fig. 2

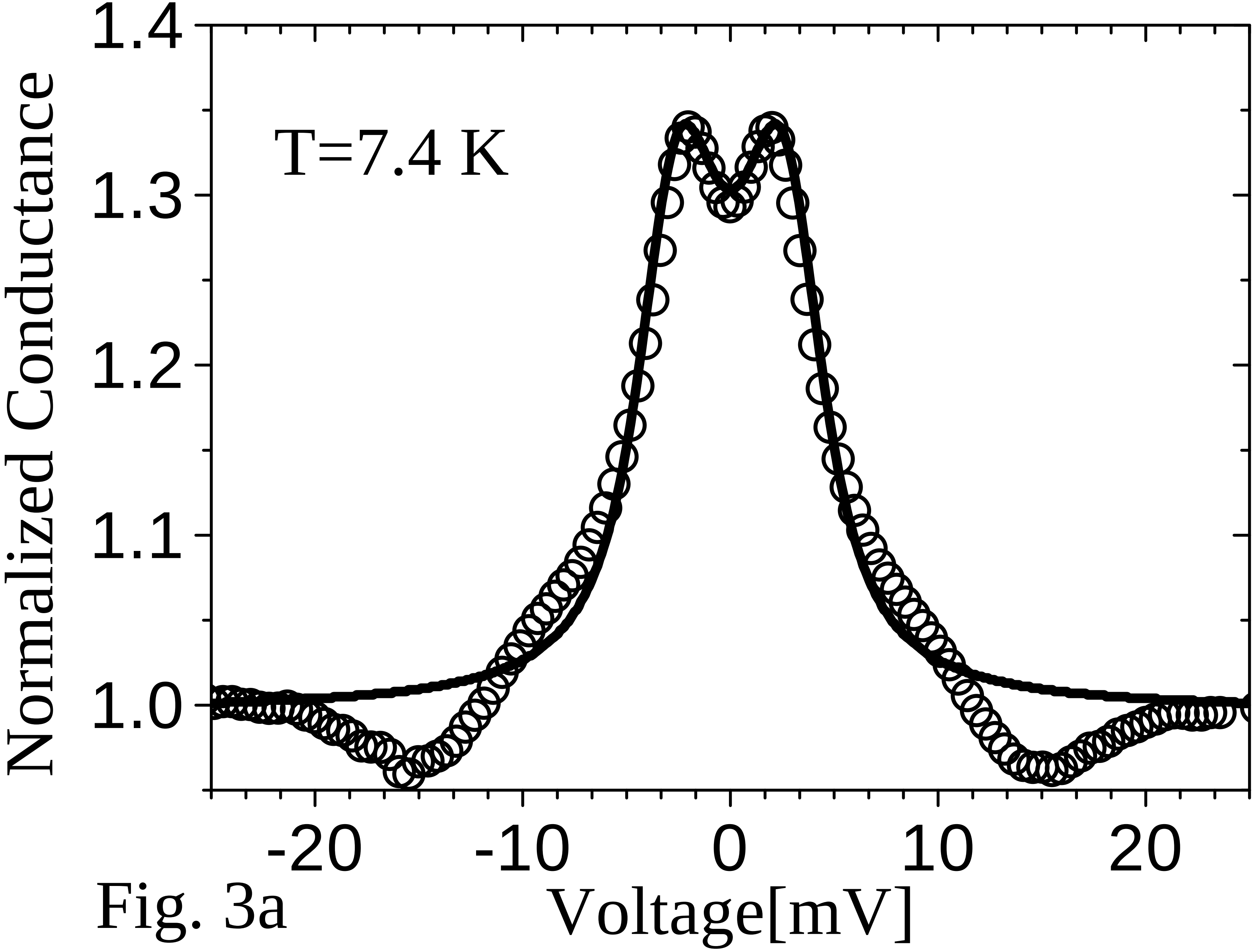
Fig. 3a

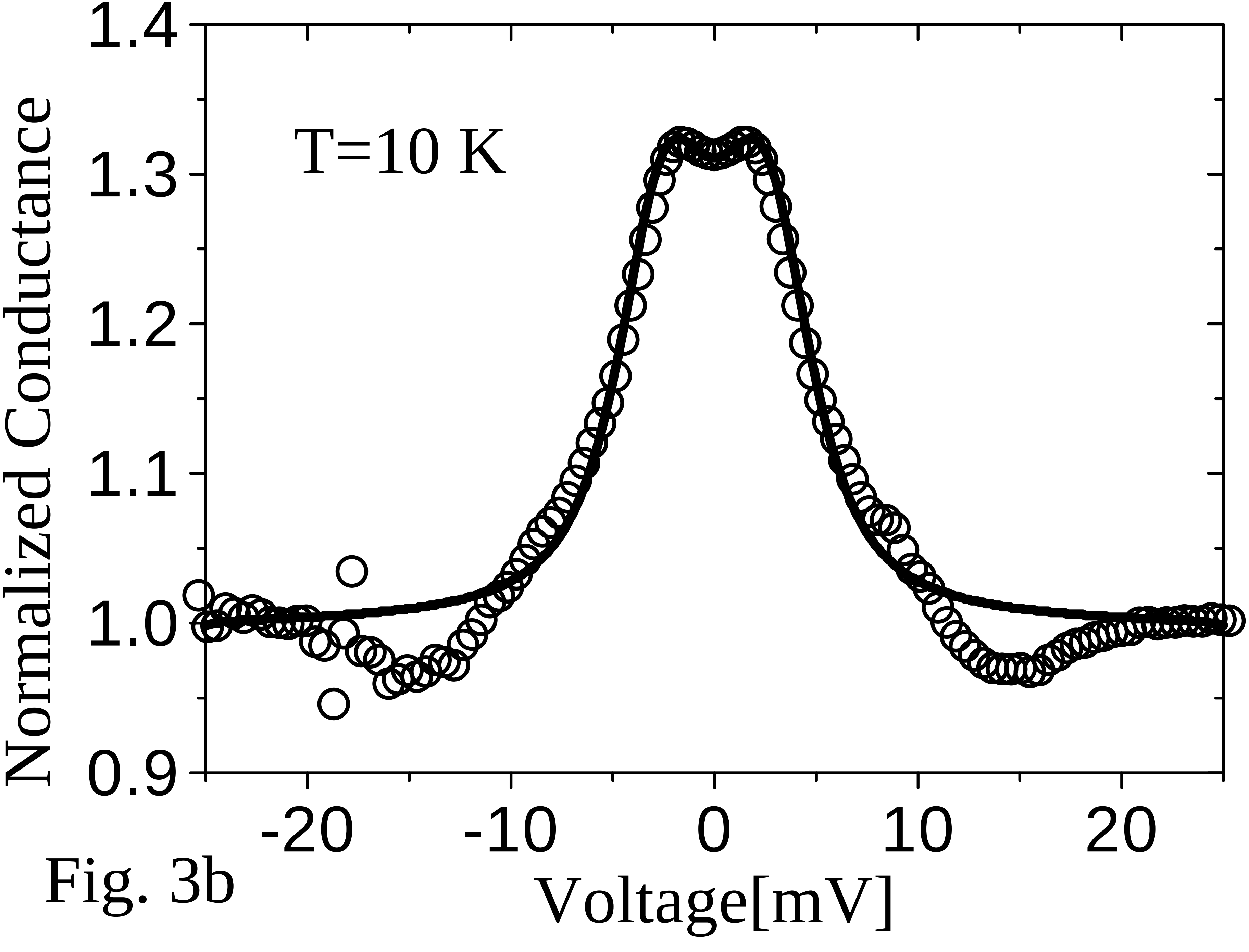
Fig. 3b

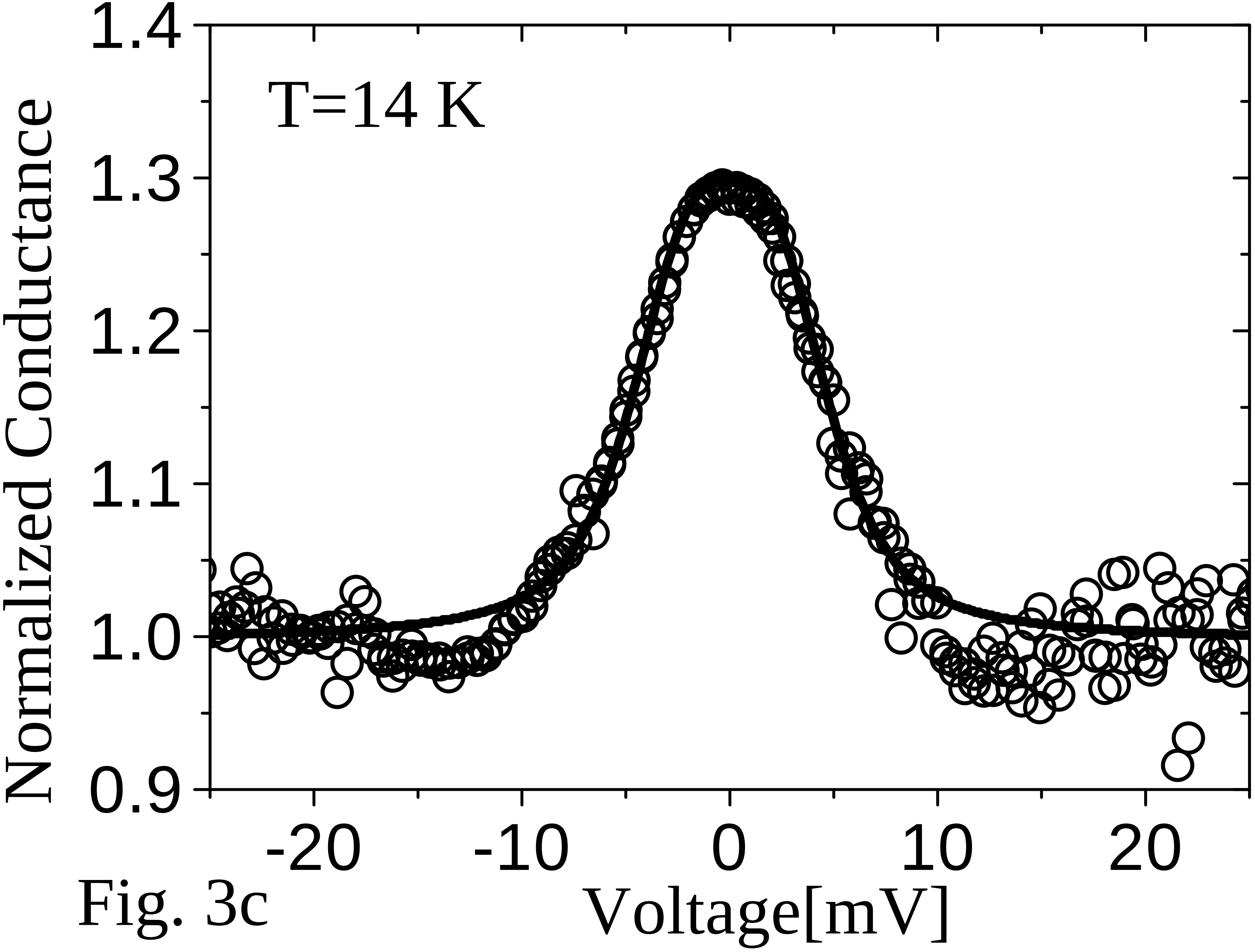
Fig. 3c

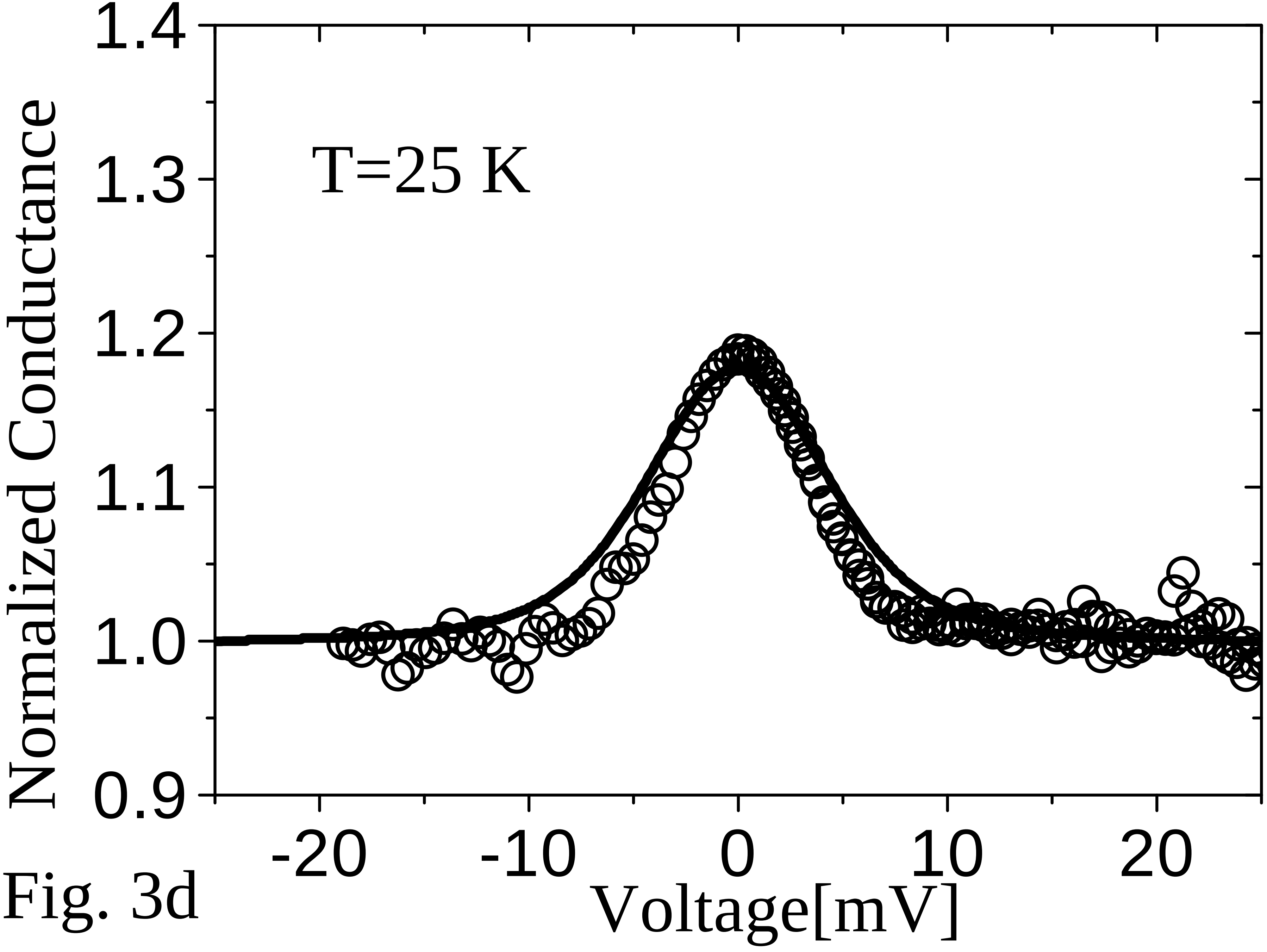

Fig. 3d

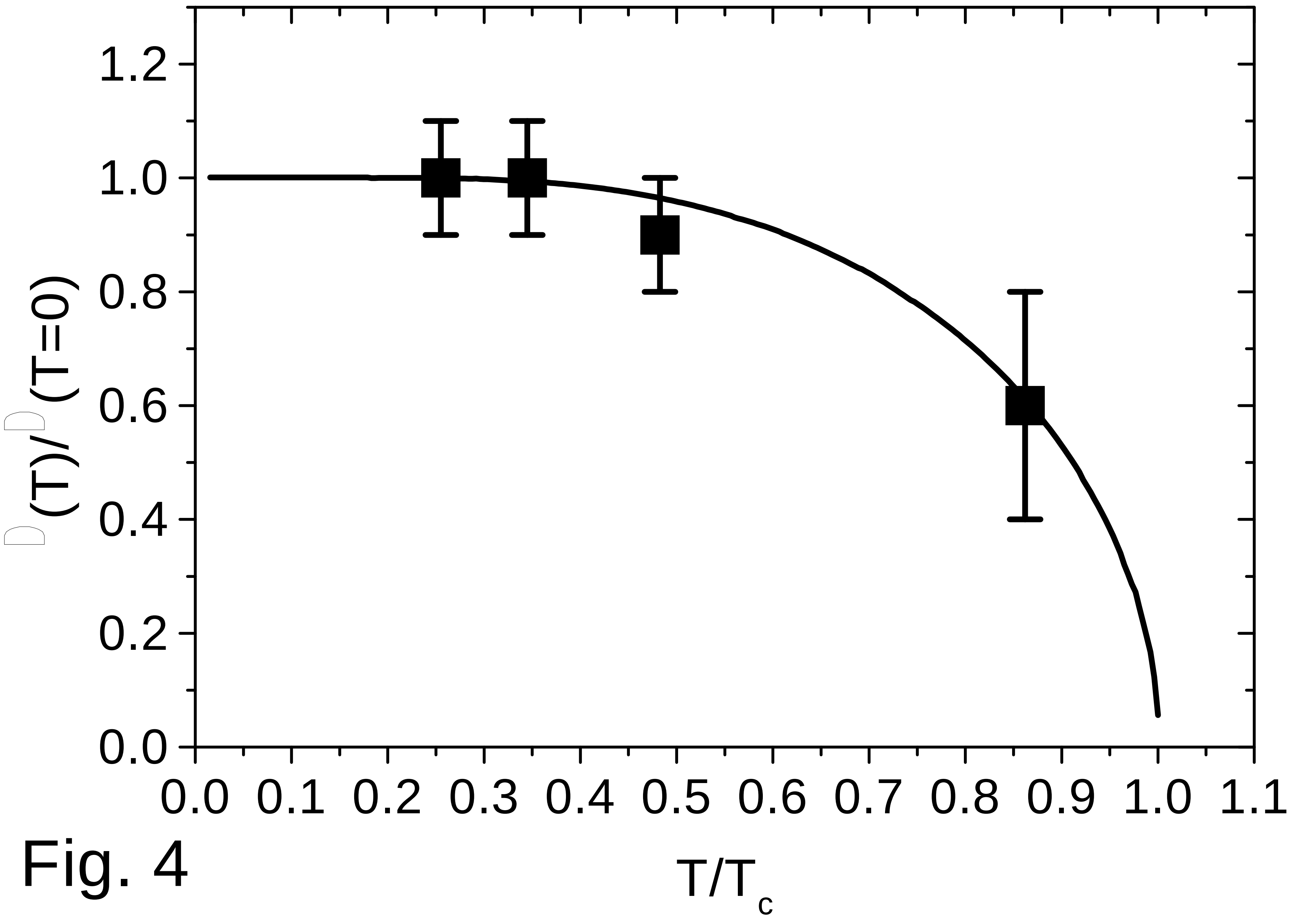

Fig. 4